\documentclass{elsart}
\usepackage[dvips]{graphicx}
\usepackage{natbib}
\usepackage{amssymb}
\begin{document}
\begin{frontmatter}
\title{Axion detection in the Micromaser}
\author{M.~L.~Jones, G.~J.~Wilkes \and B.~T.~H.~Varcoe}
\address{Department of Physics and Astronomy, University of Sussex,
Brighton, East Sussex, BN1 9QH}
\begin{abstract}
We  report  on a  scheme  for  highly  efficient detection  of  single
microwave  photons,  with  application  to  detecting  certain  exotic
particles such as the axion.  This scheme utilises an experiment known
as the micromaser and the phenomenon of trapping states to amplify the
signal and produce a cascade  of detector counts in a field ionization
detector.   The cascade  provides a  time-resolved signal  marking the
presence of a single photon. For experimentally achievable parameters,
the detection efficiency exceeds 90\%.
\end{abstract}
\begin{keyword}
Micromaser; Rydberg atoms; Cavity QED; Axion; Photon detection; Trapping
states; Quantum trajectories
\end{keyword}
\end{frontmatter}

Reliable rare-event detection is vital in many fields of physics,
especially in particle physics where there is a real need to observe the existence of rare or weakly
interacting particles. One such particle that has
received much attention is the axion \cite{raffelt}, which is predicted
to have a mass equivalent to a microwave photon. While it is very difficult to detect the axion directly schemes do exist for converting these particles into microwave photons via the Primakoff process \cite{sikivie, asztalos} where the problem becomes one of detecting the subsequent microwave photon. The use of Rydberg atoms in single microwave photon detectors \cite{matsuki96,rydbergaxion} is an attractive prospect for this purpose since it is experimentally easier to detect the effect of the photons on atoms than the photons themselves. The typical method for detecting whether a Rydberg atom
has absorbed or emitted a microwave photon is using ``state selective
field ionization'' \cite{gallagher}, where the atoms are subjected to
a varying electric field that ionizes them in a position dependent upon
their state. Detecting a single atom in an altered state reveals the presence of a photon. However, the nature of these detectors is such that a number of effects will produce detector clicks that have nothing to do with a signal and are unavoidable in practice. This has important practical implications when trying to resolve a single
detection event \cite{rydbergaxion} in the presence of background thermal noise. 

The scheme proposed in this paper produces a signal that can be discriminated from single clicks arising from random clicks unconnected with a signal, as the arrival of a single photon triggers a cascade of detector clicks. The microwave photon detection efficiency approaches 100\% and is also inherently more robust to dark counts (counts in the absence of signal), missed counts (due to finite detector efficiency) and mis-counts (arising from detector cross talk). This method of detection is based on a quantum mechanical effect called zero photon population trapping in the micromaser
\cite{weidinger99,varcoe00,varcoe04}. The micromaser plays the role of an ultrasensitive microwave detector for axions created in an ancillary conversion chamber \cite{rydbergaxion,improvedRF}. Mode matching between the two cavities would provide efficient coupling to any type of axion conversion chamber and therefore we will not comment on the structure of the conversion chamber itself. We will instead concentrate this discussion on the structure of the detector itself and describe how it achieves a high detection efficiency.  

The micromaser \cite{raithel94} is a cavity QED experiment in which we
use superconducting microwave cavities with a $Q$-factors up to $5\times
10^{10}$ (single photon lifetime of around $0.4s$), through which we
pass a sequence of two level atoms that interact one at a time with a
single mode of the cavity. The atoms are prepared initially in their
excited state and the transition between the two states is resonant with
the cavity. The atoms are very strongly coupled to the cavity mode,
therefore allowing a maser field to be produced with only one atom
passing through the cavity at a time. While the microwave photons
themselves cannot be detected, the atoms can, so we are able to derive a
great deal of information about the field from the atoms emerging from
the cavity. The micromaser has already been used to observe the
appearance of single quanta and as a triggerable source of single
microwave quanta \cite{brattke} and can be used as a microwave photon detector via population
trapping states. They occur when the cavity state is trapped and
emission of the incident atoms is forbidden until the arrival (by some
other means) of a single photon, causing a cascade of emission
events. Thus the single photon is massively amplified by the micromaser.
This occurs regardless of the atomic pump rate, which can be up to
several thousand atoms per second. When a single photon enters the
cavity the conditions change from destructive to constructive
interference and emission probability can rise to nearly 100\%. Thus the
rate of emission events can go from zero to thousands per second on the
arrival of one photon.

\begin{figure}[ht] \begin{center}
\includegraphics[width=0.8\textwidth]{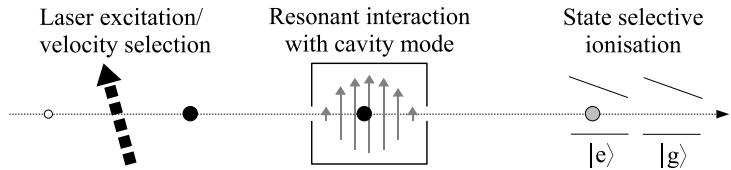} 

\caption{Schematic of operation of the micromaser. Ground state rubidium
atoms (small unfilled circle) exit the oven with thermal velocities. A
detuned angled laser excitation region excites a particular velocity
class to the $63P_{3/2}$ Rydberg state (black circles). The transition
between this and the $61D_{5/2}$ state is resonant with a single mode in
a superconducting microwave cavity and interacts coherently with it. The
atomic state is recorded by state selective field ionization detectors 
upon exiting the cavity.}

\label{fig:micromaser} 
\end{center} 
\end{figure}

Figure \ref{fig:micromaser} shows the experimental operation of the
micromaser. Rubidium atoms are emitted from an oven in their ground
state (unfilled circles) in a highly collimated beam. A laser excites
these atoms to the $63P_{3/2}$ Rydberg state (black circles), which acts
as the upper level $\vert e\rangle$ of what is effectively a two level
system. In this case, the lower level $\vert g\rangle$ is the
$61D_{5/2}$ Rydberg state, separated by 21.5GHz. However, Rydberg
transitions are closely spaced and span frequencies from 10--120GHz,
allowing us to search anywhere in a mass range of approximately
40--500$\mu$eV. The excitation laser is angled with respect to the
atomic beam to allow velocity selection via Doppler detuning. A typical
velocity resolution of 0.5\% is achievable with current techniques. The
excited atoms enter the high-$Q$ superconducting cavity and interact
resonantly with a single mode (typically the TE$_{121}$ mode) of the
resonator. This interaction is (to a very good approximation) described
by the Jaynes-Cummings Hamiltonian \cite{jaynescummings}

\begin{equation}\label{eq:jaynescummings}
\hat{H}=\hbar\omega_{0}\hat{\sigma}_{z}+\hbar\omega\left(\hat{a}^{
\dagger}\hat{a}+\frac{1}{2}\right)+\hbar
g\left(\hat{a}^{\dagger}\hat{\sigma}_{+}
+\hat{a}\hat{\sigma}_{-}\right)
\end{equation}

where $\omega_{0}$ and $\omega$ are the atomic transition and field mode
frequencies respectively, $\hat{\sigma}_{z}$ is the atomic projection
operator, $\hat{\sigma}_{\pm}$ are the atomic raising and lowering
operators, $g$ is the coupling strength ($\approx 40\mathrm{krads^{-1}}$
in the micromaser) between atom and field, $\hat{a}$ and
$\hat{a}^{\dagger}$ are the photon annihilation and creation operators.
This is one of the simplest Hamiltonians in quantum optics, describing
the interaction of a single two-level atom with a single field mode.
There are no loss mechanisms in this ideal model, but since the
interaction time is 3--4 orders of magnitude smaller than the time
scales for losses (the micromaser cavity has a $Q$-factor of up to
$5\times 10^{10}$), this is an excellent approximation.

During the interaction, the system undergoes Rabi oscillations between
the states $\vert e\rangle$ and $\vert g\rangle$. After exiting the
cavity, the state of the atom is measured by state selective field
ionization, giving us information about the field. The probability that
an atom emits a photon into the cavity is given by 
\begin{equation}
P_{\mathrm{emit}}=\sin^{2}\left(g \tau\sqrt{n+1}\right)\label{eq:prob}
\end{equation} 
where $\tau$ is the interaction time and $n$ is the
number of photons already in the cavity. By tuning the atomic velocity
correctly, we can reduce this probability to zero, which can be
understood as the system undergoing an integer number of Rabi
oscillations, 
\begin{equation}
\tau=\frac{k\pi}{g\sqrt{n+1}}\label{eq:trapping} 
\end{equation} 
where $k$ is an integer. If equation \ref{eq:trapping} is fulfilled,
then the cavity field is trapped with $n$ photons and has zero
probability of progressing to $n+1$, which is possible even for $n=0$
(known as the vacuum trapping state). Such trapping states have been
experimentally observed \cite{weidinger99, varcoe00}. By starting in the
vacuum state (which is ensured by cooling the cavity to around 40mK) and
appropriately tuning the velocity (and hence the interaction time)
according to the value for the vacuum trapping state, emission is
forbidden and atoms are unable to emit photons. However, if $n=1$, the
probability of emission at the same velocity is close to 93\%. Thus the
addition of one photon has a dramatic effect on emission probability and
the count rate in the $\vert g\rangle$ detector goes from zero to a
detectably high number, indicating the arrival of a single photon in the
cavity. Subsequent emission events change the emission probability via
equation \ref{eq:prob}. These probabilities mean that the arrival of a
single photon in the cavity by any mechanism causes the system to jump
past the vacuum trapping state, giving a cascade of $\vert g\rangle$
detector counts. This is the signature of a single photon arriving in
the cavity.

The proposed experimental configuration is similar to that in
\cite{rydbergaxion}, with the detection cavity replaced by the
micromaser system presented here.

To examine the effectiveness of this scheme, we have employed a Quantum
Trajectory Analysis method to simulate detection records of the system.
This technique involves
stochastically evolving a wavefunction using a combination of a
non-Hermitian Hamiltonian and a set of ``jump operators''. Quantum
trajectory analysis is often used to calculate an approximation to the
density matrix of a system, but here individual trajectories are used to
simulate detection records from the experiment.
The implementation of the method here is based upon that in
\cite{cresser96}. The non-Hermitian effective Hamiltonian is given by
\begin{equation}
\hat{H}_{\mathrm{eff}}=-\frac{1}{2}i\hbar\gamma
\left[\left(n_{t}+1\right)\hat{a}^{\dagger}\hat{a}+n_{t}\hat{a}\hat{a}^{
\dagger}\right]-\frac{1}{2}i\hbar
R+\hbar\omega\hat{a}^{\dagger}\hat{a}
\end{equation}
where $\gamma$ is the cavity decay constant, $n_{t}$ is the
thermal photon number and $R$ is the rate at which atoms pass through
the
cavity.

The particular set of jump operators used are listed below:
\begin{equation}
\hat{C}_{-1}=\sqrt{\gamma \left(n_{t}+1\right)}\hat{a}
\end{equation}
is the operator that represents a photon being lost to the
reservoir,
\begin{equation}
\hat{C}_{0}=\sqrt{R}\cos\left(g\tau\sqrt{n+1}\right)
\end{equation}
represents an atom traversing the cavity and
exiting in its original excited state,
\begin{equation}
\hat{C}_{1}=\sqrt{R}\frac{\sin\left(g\tau\sqrt{n}\right)}{\sqrt{n}}
\hat{a}^{\dagger}
\end{equation}
is the operator representing an atom introducing a photon into the field
that contains $n$ photons, and
\begin{equation}
\hat{C}_{2}=\sqrt{\gamma n_{t}}\hat{a}^{\dagger}
\end{equation}
is the operator representing a photon being gained from the reservoir.
However, since every operator maps pure states onto pure states, and for
our purposes we always begin with the (pure) vacuum state, then we can
reduce the dynamics simply to jumps occurring stochastically and the
wavefunction remaining unchanged in between.

\begin{figure}[ht]
\begin{center}
\includegraphics[width=0.8\textwidth]{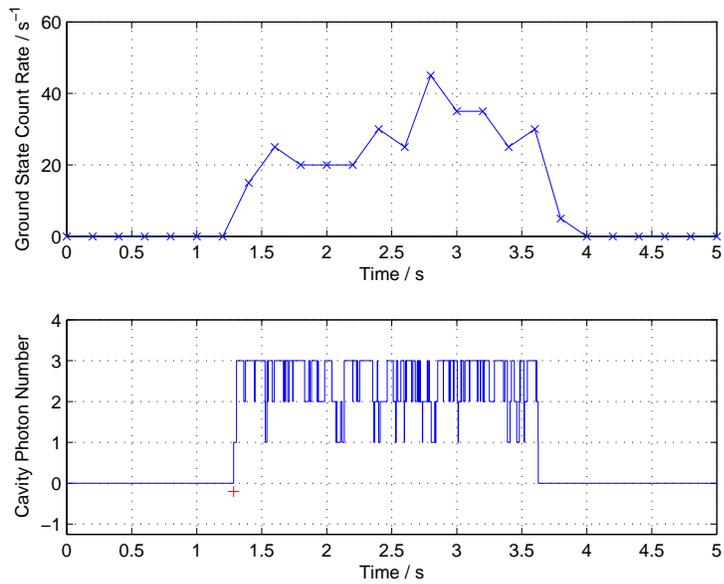}
\caption{A pair of graphs showing how the change in the cavity
photon number effects the ground
state count rate for ideal detectors. Notice that a very small
change in the cavity photon number can produce a very large change in
count rate.}\label{fig:results1}
\end{center}
\end{figure}

A quantum trajectory simulation of the micromaser operating in the
vacuum trapping state was performed, in which the ground state detector
count (the rate of occurence of jump $\hat{C}_2$) was monitored while
single photons were added to the cavity at random times. Figure
\ref{fig:results1} shows an example trajectory for the ideal case, with
no deviation from perfect operating conditions, in order to illustrate
the principle of the operation of the detector. It shows how the field
evolves inside the cavity, along with the detector clicks we see when
probing the atoms.

Here we see that, for sufficiently high atomic pump rates, once the
vacuum trapping state is passed then the field very quickly reaches
three photons, which also gives a zero emission probability and becomes
trapped again (this is the $n=3$, $k=2$ trapping state,
eq.~\ref{eq:trapping}). The field then proceeds to rapidly oscillate
between this and the two and one photon states (due to decay from the
cavity), giving rise to the high count rate. We see that adding just one
photon at around $t=1.25s$ produces a detector count rate of up to
around 45 counts per second, which is easily detectable, even with
imperfect detectors. Notice that, for a typical microwave frequency of
around 21.5GHz, as used in current micromaser experiments, this amounts
to near perfect detection of an energy change of less than $90\mu eV$.
                                   
\begin{figure}[ht] \begin{center}
\includegraphics[width=0.8\textwidth]{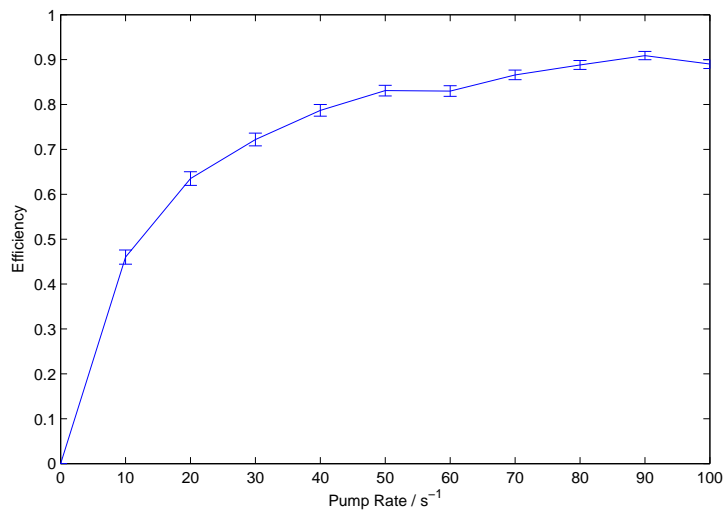} 

\caption{A graph showing how the detection efficiency increases for
increasing values of atomic flux $R$. The error bars indicate the
statistical spread of simulated results.
}
\label{fig:efficiency}
\end{center} 
\end{figure} 
Figure \ref{fig:efficiency} shows how the detection efficiency depends
upon the pump rate $R$. This simulation was performed by introducing a
single photon into the cavity at a random time, and if a ground state
count rate above a threshold of 10Hz was achieved within a set interval,
then a successful detection was said to have occurred. This process was
repeated 1000 times for each value of $R$ to give an average detector
efficiency. The velocity spread and temperature were both set to zero in
this case.

\begin{figure}[ht]
\begin{center}
\includegraphics[width=0.8\textwidth]{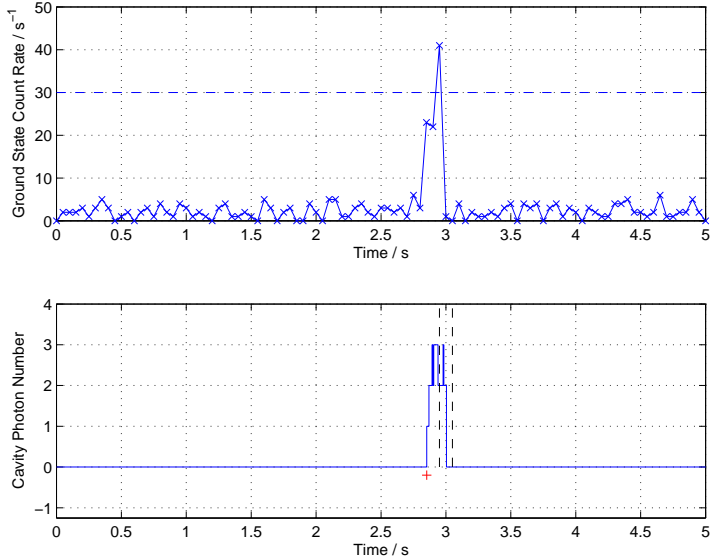}
\caption{Plots to show how the proposed threshold system would work. 
Once the count rate exceeds the predefined threshold, a detection event 
is recorded and the cavity field is allowed to relax back to the vacuum
state by switching off the excitation laser or applying a $\pi-$pulse to 
the excited atoms to pump the cavity with ground state atoms.}
\label{fig:threshold}
\end{center}
\end{figure}

Figure \ref{fig:threshold} displays the threshold operation (in this
case with an artificially high threshold of 30 counts / second for
illustrative purposes). To prepare for the next count period, the field
is then allowed to relax back to the vacuum state, either via free decay
of the field, or more quickly by pumping with ground state atoms. This
period of dead time is shown in the lower plot of
fig.~\ref{fig:threshold} delineated by verical dashed lines. Applying a
$\pi-$pulse to the incoming atoms in state $\vert e\rangle$ evolves
their state to $\vert g\rangle$, which allows for faster pump-down to
the vacuum state. Additionally, by relaxing the velocity selection, for
example by using a perpendicular excitation scheme, the pump-down rate
to the vacuum state can be further enhanced. This method would allow
detector dead-times significantly shorter than simple cavity decay alone
would permit.

To be more realistic, however, it is possible to include a number of
departures from the ideal conditions in the quantum trajectory method to
investigate the limits imposed on the system by these factors. For
example, a more complete model of this system includes practical
limitations of the system. \emph{Dark Counts} are caused by detector
clicks occurring by means other than ionization of the rubidium atoms
(for example simple thermal excitation in the detector or cosmic rays
passing through the detector), leading to non-zero count rates when no
atoms are present, giving a Poisson distributed background level. High
quality electron multipliers reduce this rate to around 3 counts per
second or less. \emph{Missed Counts} arise when an atom is not ionized
at all in the field ionization region, or when the liberated electron
does not reach the electron multiplier. \emph{Detector Crosstalk} occurs
when an atom is ionized at the wrong detector, leading to errors in the
statistics of the detected atoms.

All of these errors are incorporated into the model by means of setting
the detector efficiencies $\eta_g<1$ and $\eta_e<1$ for the ground and
excited state detectors respectively and adding a random background
generated with a poissonian distribution centred at $r_{b}$
counts/second to simulate the dark counts and crosstalk.

Other errors in the system arise from the departure from ideal operating
conditions of the micromaser. The idealised model assumes that there is
no spread in interaction time $\tau$, the coupling parameter $g$ is
constant and that there is never more than one atom in the cavity at any
time. In practice, we find that, due to the linewidths of the velocity
selecting laser and atomic transition, the interaction time has a
non-zero spread. Mechanical vibrations in the system may also cause
variations in the parameter $g$. Hence we replace $g$ and $\tau$ with
$\phi=g\tau$, drawn from a normal distribution centred at $\phi_{0}=\pi$
with spread $\Delta\phi$ to represent these effects. Perhaps the major
source of error, however, is the occurrence of multi-atom events. When
there is more than one atom in the cavity, equation
\ref{eq:jaynescummings} no longer holds, and the more complicated
interaction has a high probability of breaking the trapping state
barrier and causing an erroneous detection event. The probability of an
atom contributing to a single atom event is given by $P=e^{-2R\tau}$,
which gives a maximum rate of $62s^{-1}$ for a $99\%$ probability of one
atom events at our vacuum trapping state. This effect is easily included
in the simulation by monitoring the time between incident atoms. If two
or more are present in the cavity, then the trapping state is broken by
the addition of one or two photons.

\begin{figure}
\begin{center}
\includegraphics[width=0.8\textwidth]{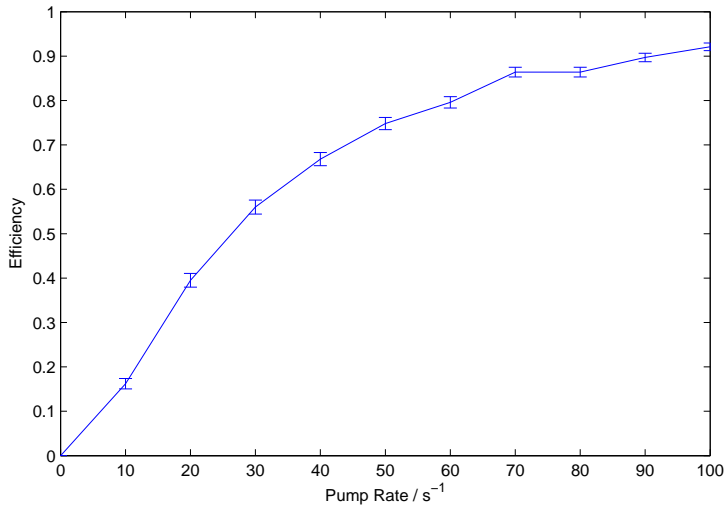}
\caption{Plot of detector efficiency for different values of $R$, with
associated error bars. 
The parameter values are: $r_{b}=2$, $\Delta\phi=0.5\%$, $\eta=0.7$, 
threshold = 10.}\label{fig:finalEff}
\end{center}
\end{figure}
\begin{figure}
\begin{center}
\includegraphics[width=0.8\textwidth]{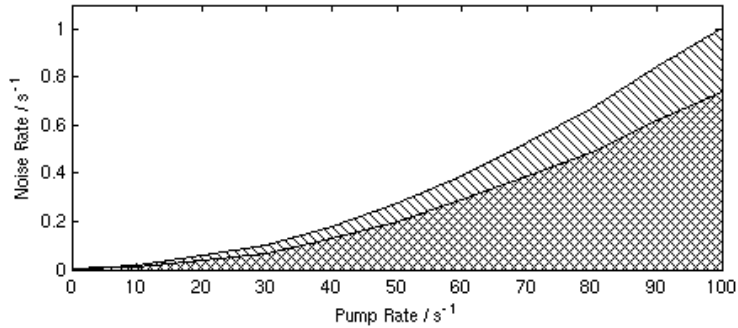}
\caption{Plot showing the composition of background counts. The lower 
portion represents the noise rate due to multi-atom events and the upper 
portion that of the finite velocity spread.}\label{fig:finalNoise}
\end{center}
\end{figure}

We now see that the detection efficiency increases with increasing pump
rate, but that the background counts also increase, due to the higher
probability of two-atom events and outliers in the velocity distribution
that disrupt the trapping state. This clearly affects the signal to
noise ratio, and by plugging in real experimental numbers for our errors
we can easily predict the optimum operating conditions to maximise our
signal to noise ratio.

Figure \ref{fig:finalNoise} shows that, for experimentally realistic
parameters, multi-atom effects are the largest single source of noise.
The effect of these multi-atom events can be reduced by increasing the
coupling $g$, and hence reducing $\tau$, or altering the distribution of
incoming atoms. In the limit of uniform atomic spacing, for example, we
can in principle achieve an upper limit of $R=\tau^{-1}\approx 10^{4}$
(a promising method of altering the distribution is currently being
investigated).

In this paper we have shown that it is possible to massively amplify the
characteristic signal of a single microwave photon to a level where it
is easily measurable with current detector technology. Furthermore, the
theoretical model presented here can be used to decide the particular
operating parameters for optimal performance. 

\ack We would like to acknowledge the support of the EPSRC network
UKCAN. The work is being funded by EPSRC and MJ and GW are being funded
by EPSRC DTA studentships. 
\bibliographystyle{elsart-num}
\bibliography{Bibliography} 

\begin{thebibliography}{10}
\expandafter\ifx\csname url\endcsname\relax
  \def\url#1{\texttt{#1}}\fi
\expandafter\ifx\csname urlprefix\endcsname\relax\def\urlprefix{URL }\fi

\bibitem{raffelt}
G.~Raffelt, Axions, Space Science Reviews 100 (2002) 153--158.

\bibitem{sikivie}
P.~Sikivie, Experimental tests of the ``invisible'' axion, Phys.~Rev.~Lett. 51
  (1983) 1415--1417.

\bibitem{asztalos}
S.~Asztalos, E.~Daw, H.~Peng, L.~Rosenberg, C.~Hagmann, D.~Kinion, W.~Stoeffl,
  K.~{van Bibber}, P.~Sikivie, N.~Sullivan, D.~Tanner, F.~Nezrick, M.~Turner,
  D.~Moltz, J.~Powell, M.~Andr\'e, J.~Clarke, M.~M\"uck, R.~Bradley,
  Large-scale microwave cavity search for dark-matter axions, Phys.~Rev.~D.
  64~(092003).

\bibitem{matsuki96}
I.~Ogawa, S.~Matsuki, K.~Yamamoto, Interaction of cosmic axions with rydberg
  atoms in resonant cavities via the primakoff process, Phys.~Rev.~D. 53~(4).

\bibitem{rydbergaxion}
M.~Tada, Y.~Kishimoto, K.~Kominato, M.~Shibata, H.~Funahashi, K.~Yamamoto,
  A.~Masaike, S.~Matsuki, Carrack {II} -- a new large-scale experiment to
  search for axions with rydberg-atom cavity detector, Nuc.~Phys.~B
  (Proc.~Supp.) 72 (1999) 164--168.

\bibitem{gallagher}
T.~F. Gallagher, Rydberg atoms, Cambridge University Press, 1994.

\bibitem{weidinger99}
M.~Weidinger, B.~T.~H. Varcoe, R.~Heerlein, H.~Walther, Trapping states in the
  micromaser, Phys.~Rev.~Lett. 81 (1999) 5784--5787.

\bibitem{varcoe00}
B.~T.~H. Varcoe, S.~Brattke, M.~Weidinger, H.~Walther, Preparing pure photon
  number states of the radiation field, Nature 403~(6771) (2000) 743--746.

\bibitem{varcoe04}
B.~T.~H. Varcoe, S.~Brattke, H.~Walther, The creation and detection of
  arbitrary photon number states using cavity {QED}, New Journal of Physics
  6~(97).

\bibitem{improvedRF}
S.~J. Asztalos, R.~F. Bradley, L.~Duffy, C.~Hagmann, D.~Kinion, D.~M. Moltz,
  L.~J. Rosenberg, P.~Sikivie, W.~Stoeffl, N.~S. Sullivan, D.~B. Tanner, K.~van
  Bibber, D.~B. Yu, Improved rf cavity search for halo axions, Phys.~Rev.~D
  69~(011101(R)).

\bibitem{raithel94}
G.~Raithel, C.~Wagner, H.~Walther, L.~M. Narducci, M.~O. Scully, The
  micromaser: a proving ground for quantum physics, in: P.~R. Berman (Ed.),
  Cavity Quantum Electrodynamics, Advances in atomic, molecular and optical
  physics, Academic Press, New York, 1994.

\bibitem{brattke}
S.~Brattke, B.~T.~H. Varcoe, H.~Walther, Generation of photon number states on
  demand via cavity quantum electrodynamics, Phys.~Rev.~Lett. 86 (2001)
  3534--3537.

\bibitem{jaynescummings}
E.~T. Jaynes, F.~W. Cummings, Comparison of quantum and semiclassical radiation
  theories with application to the beam maser, Proc.~IEEE. 51 (1963) 89.

\bibitem{cresser96}
J.~D. Cresser, S.~M. Pickles, A quantum trajectory analysis of the one-atom
  micromaser, Quantum Semiclass.~Opt. 8~(1) (1996) 73--104.

\end{thebibliography}
\end{document}